\documentclass{aa}  

\usepackage{graphicx}
\usepackage{txfonts}
\usepackage{threeparttable}
\usepackage{graphicx}
\usepackage{amssymb}
\usepackage{amsmath}
\usepackage{algorithm} 
\usepackage{algpseudocode}
\usepackage[normalem]{ulem}

\newcommand{\RM} [1]{\mathrm{#1}}

\newcommand{\EQ}[1]{Equation~(\ref{eq:#1})}
\newcommand{\FIG}[1]{Fig.~\ref{fig:#1}}

\newcommand{\SEC}[1]{Section~\ref{sec:#1}}
\newcommand{\APP}[1]{APPENDIX~\ref{sec:#1}}

\begin{document} 

   \title{TransientX: A high performance single pulse search package}


   \author{Yunpeng Men\inst{1}
          \and
          Ewan Barr\inst{1}
          }

   \institute{Max-Planck-Institut f\"{u}r Radioastronomie, Auf dem H\"{u}gel 69, D-53121 Bonn, Germany\\
              \email{ypmen@mpifr-bonn.mpg.de}
             }

   \date{Received XX XX, XXXX; accepted XX XX, XXXX}

 
  \abstract
   {Radio interferometers composed of a large array of small antennas posses large fields of view, coupled with high sensitivities. For example, the Karoo Array Telescope (MeerKAT), achieves a gain of up to 2.8 K/Jy across its $>1\,\mathrm{deg}^2$ field of view. This capability significantly enhances the survey speed for pulsars and fast transients. Nevertheless, this also introduces challenges related to the high data rate, reaching a few Tb/s for MeerKAT, and substantial computing power requirements.}
   {To handle the large data rate of surveys, we have developed a high-performance single-pulse search software called "TransientX". This software integrates multiple processes into one pipeline, which includes radio frequency interference mitigation, de-dispersion, matched filtering, clustering, and candidate plotting.}
   {In {\sc{TransientX}}, we have developed an efficient CPU-based de-dispersion implementation using the sub-band de-dispersion algorithm. Additionally, {\sc{TransientX}} employs the density-based spatial clustering of applications with noise (DBSCAN) algorithm to eliminate duplicate candidates, utilizing an efficient implementation based on the kd-tree data structure. We also calculate the signal-to-noise ratio loss resulting from dispersion measure, boxcar width, spectral index and pulse shape mismatches. Remarkably, we find that the signal-to-noise ratio loss resulting from the mismatch between a boxcar-shaped template and a Gaussian-shaped pulse with scattering remains relatively small, at approximately 9\%, even when the scattering timescale is 10 times that of the pulse width. Additionally, the S/N decrease resulting from the spectra index mismatch becomes significant with multi-octave receivers.}
   {We have benchmarked the individual processes, including de-dispersion, matched filtering, and clustering. Our de-dispersion implementation can be executed in real-time using a single CPU core on data with 4096 dispersion measure (DM) trials, which consist of 4096 channels and have a time resolution of 153 microseconds. Overall, {\sc{TransientX}} offers the capability for efficient CPU-only real-time single pulse searching.}
   {}


   \maketitle
%

\section{Introduction}
The single-pulse search technique has been employed in the search for pulsars since the first pulsar discovery \citep[e.g.][]{Hewish1968Nat, Large1968Nat}. The timescale for a single-pulse search can vary widely, ranging from microseconds to tens of seconds \citep[e.g.][]{Snelders2023NatAs, HurleyWalker2023Nat}. This technique proves particularly useful in cases where pulsar signals exhibit nulling emissions in most of their periods \citep[e.g.][]{Keane2011Obs}, as it can significantly enhance the signal-to-noise ratio (S/N). Remarkably, single-pulse searches have led to the discovery of a new class of pulsars known as rotating radio transients (RRATs). RRATs exhibit extreme nulling behavior in comparison to typical pulsars  \citep[e.g.][]{McLaughlin2006Nat}. Furthermore, this technique has played a pivotal role in the discovery of a new class of astronomical objects known as fast radio bursts (FRBs). FRBs are characterised by their bright, short-duration radio bursts, the origin of which remains unknown \citep[e.g.][]{CHIME2021ApJS}.

There are several widely used open-source single-pulse search packages, including {\sc{PRESTO}\footnote{https://github.com/scottransom/presto.git}} \citep{Ransom2011ascl} and {\sc{HEIMDALL}\footnote{https://sourceforge.net/projects/heimdall-astro/}}. {\sc{PRESTO}} is a CPU-based toolset that comprises individual programs designed for various tasks, including radio frequency interference (RFI) mitigation, de-dispersion, and single-pulse search \citep{Ransom2011ascl}. On the other hand, {\sc{HEIMDALL}} is a GPU-accelerated transient detection pipeline \citep{Barsdell2012MNRAS}. It leverages the computational power of graphics processing units (GPUs) for faster processing. Another CPU-based single-pulse search pipeline is BEAR, which integrates multiple processes into a single program. These processes encompass RFI mitigation, de-dispersion, clustering, and candidate plotting \citep{Men2019MNRAS}.

In this study, we introduce a novel high-performance single-pulse search software named {\sc{TransientX}\footnote{https://github.com/ypmen/TransientX}}, designed as a data block-based pipeline, which handles the data in successive segments of typically a few seconds. {\sc{TransientX}} incorporates more advanced RFI mitigation algorithms and employs a more efficient clustering algorithm, than its predecessor, {\sc{BEAR}}. Additionally, {\sc{TransientX}} boasts a highly efficient de-dispersion implementation, as demonstrated in \SEC{dedispersion}. To further enhance performance, {\sc{TransientX}} has been optimised by using AVX2 instructions and the multiprocessing library, OpenMP. Typically, {\sc{TransientX}} exhibits a significant speed improvement of approximately one order of magnitude compared to {\sc{BEAR}}. Improving performance can offer numerous benefits, such as expanding the search parameter space, saving energy and environmental costs. Furthermore, it can reduce data processing time to enable trigger observation.

In this paper, we introduce the algorithms employed in {\sc{TransientX}} in \SEC{Algorithm}. We present the benchmark results in \SEC{Benchmark}. Our discussion is provided in \SEC{Discussions}, and we summarize our conclusions in \SEC{Conclusions}.

\section{Algorithm}
\label{sec:Algorithm}

In {\sc{TransientX}}, the data are batch processed in blocks of typically a few seconds in length depending on the maximum dispersion delay. Each data block then undergoes a series of processing stages, which include normalisation, RFI mitigation, de-dispersion, matched filtering, clustering, and candidate plotting. We present the algorithms used in these processes in the following subsections.
\begin{figure}
    \centering
    \includegraphics[width=\columnwidth]{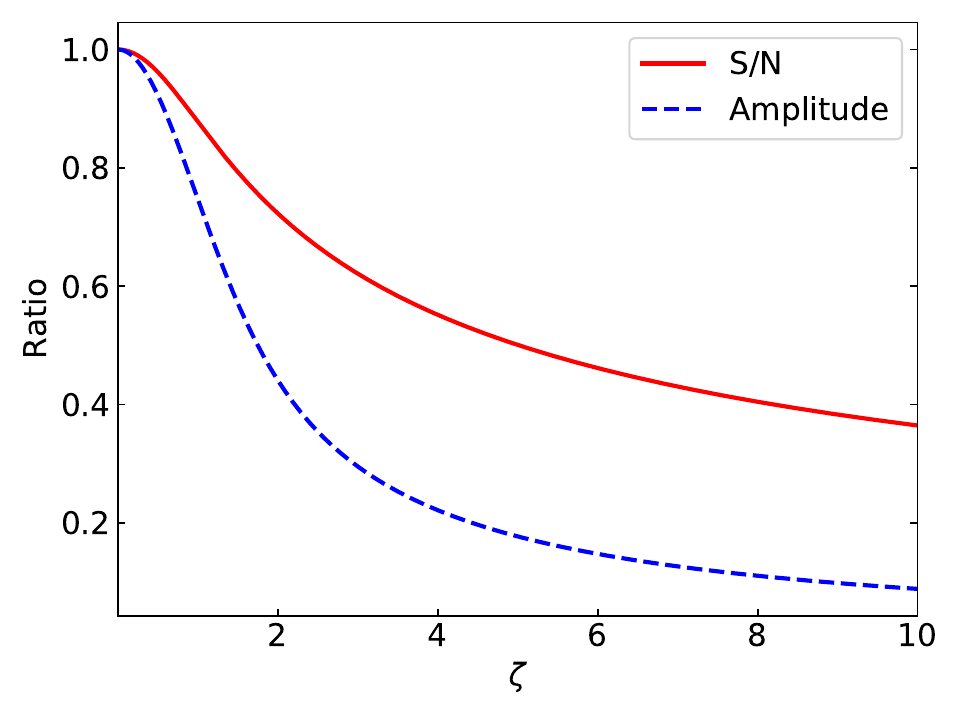}
    \caption{Relationship between the S/N or amplitude ratio and the DM mismatch factor $\zeta$ defined in \EQ{zeta}. The red solid line represents the S/N relationship, whereas the blue dashed line represents the amplitude relationship.}
    \label{fig:dm_mismatch}
\end{figure}

\begin{figure*}
    \centering
    \includegraphics[width=1.8\columnwidth]{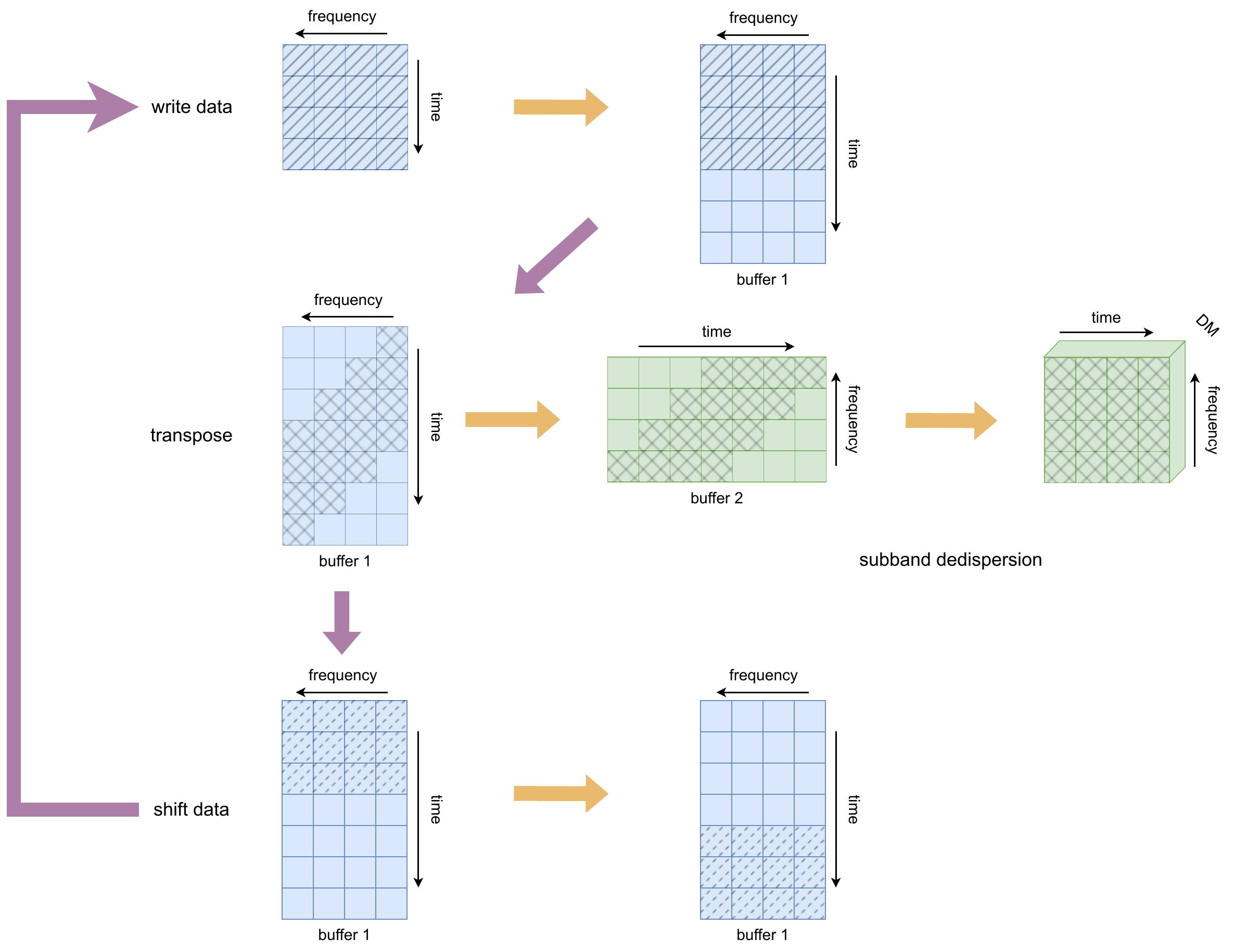}
    \caption{Dataflow diagram for de-dispersion, incorporating the stages of "write data," "transpose," "sub-band de-dispersion," and "shift data." In this diagram, green memory blocks are utilised to buffer incoming data, while blue blocks serve as buffers for transposed data, where the de-dispersion process takes place. The shaded patterns within the diagram illustrate data mapping at various steps: (1) In the "write data" step, the shaded pattern represents the positions where the input data is written. (2) In the "transpose" step, the shaded pattern illustrates how the dedispersed data is transposed in buffer 1 and buffer 2. (3) In the "subband dedispersion" step, the shaded pattern represents the dedispersion of the data. (4) In the "shift data" step, the shaded pattern illustrates the mapping of the data as it is shifted.}
    \label{fig:dedispersion_flow}
\end{figure*}

\subsection{Normalisation}
\label{sec:Normalisation}
Due to the typically frequency-dependent system response, which results in substantial variations in noise and bias levels across different frequency channels of the data, we implement data normalisation. This normalisation involves adjusting the data to have a mean of zero and unity variance within each frequency channel. Notably, the normalisation process can also mitigate the impact of outliers in frequency channels by reweighting the data with the reciprocal of its variance, thus preventing S/N drops. Additionally, it facilitates subsequent processes by providing normalised data.

\subsection{RFI mitigation}
\label{sec:RFI mitigation}

Radio frequency interference are non-astronomical signals that usually come from electronic devices, satellites, lighting, and so on. The RFI signals can lead to false positive candidates in the single-pulse search, so RFI mitigation is applied in the processing to reduce these false candidates. There are some effective RFI mitigation algorithms that have been proposed \citep[e.g.][]{Offringa2012A&A, Men2019MNRAS, Morello2022MNRAS, Men2023arXiv}. RFI signals can present as both narrowband and broad-band signals. To remove the narrowband RFI signals, {\sc{TransientX}} applies the skewness-kurtosis filter (SKF), which removes the frequency channels that are outliers in the skewness and kurtosis statistics across all frequency channels. To remove the broad-band RFI signals, {\sc{TransientX}} applies the Zero-DM Matched Filter (ZDMF), which removes the correlation components between frequency channels. Details about the algorithms can be found in \cite{Men2019MNRAS} and \cite{Men2023arXiv}.

\subsection{De-dispersion}
\label{sec:dedispersion}

Dispersion occurs when radio waves traverse through plasma, causing a delay between different frequencies. The cold plasma dispersion delay can be expressed as
\begin{equation}
     \tau_d = \frac{e^2}{2\pi m_e c} \RM{DM} \left(\frac{1}{f_1^2}-\frac{1}{f_2^2}\right)\,,
     \label{eq:dispersion_delay}
 \end{equation}
where $e,m_e,c$ represent the elementary charge, electron mass, and the speed of light in a vacuum, respectively. The dispersion measure ($\RM{DM}$) represents the column density of free electrons along the path to the source, given by
\begin{equation}
    \RM{DM} = \int n_e d l\,,
\end{equation}
where $n_e$ denotes the electron density and $l$ represents the distance to the source. De-dispersion is a critical process used to correct the delays between frequency channels, thereby enhancing the band-integrated S/N of the pulse. \cite{Cordes2003ApJ} investigated the impact of a DM mismatch on the amplitude of an individual pulse. For a Gaussian-shaped pulse with a full width at half maximum (FWHM) of $W$, the ratio of the measured peak flux $S(\delta \RM{DM})$ to the true peak flux $S$ for a DM mismatch $\delta \RM{DM}$ is given by 
\begin{equation}
    \frac{S(\delta \RM{DM})}{S} = \frac{\sqrt{\pi}}{2} \zeta^{-1} \RM{erf} \zeta \,,
\end{equation}
where
\begin{equation}
    \zeta = 6.91 \times 10^{-3} \delta \RM{DM} \frac{\Delta v_\RM{MHz}}{W_\RM{ms} v^3_\RM{GHz}}\,.
    \label{eq:zeta}
\end{equation}
$\Delta v_\RM{MHz}$ represents the bandwidth, and $v_\RM{GHz}$ is the central radio frequency in units of MHz and GHz, respectively. However, in practical applications, our concern is typically the S/N loss rather than just the decrease in amplitude. In this study, we conducted semi-analytical calculations to determine the ratio of the measured S/N to the optimal S/N, as presented in \APP{dm_mismatch}. \FIG{dm_mismatch} illustrates the comparison between the decrease in amplitude and the decrease in S/N defined in \EQ{snr}. Notably, it demonstrates that the S/N decrease is slower than the decrease in amplitude. This happens because, even though the pulse gets temporally broadened due to the DM mismatch, a wider boxcar picks it up, slowing down the decrease in S/N.

\begin{figure*}
    \centering
    \includegraphics[width=1.6\columnwidth]{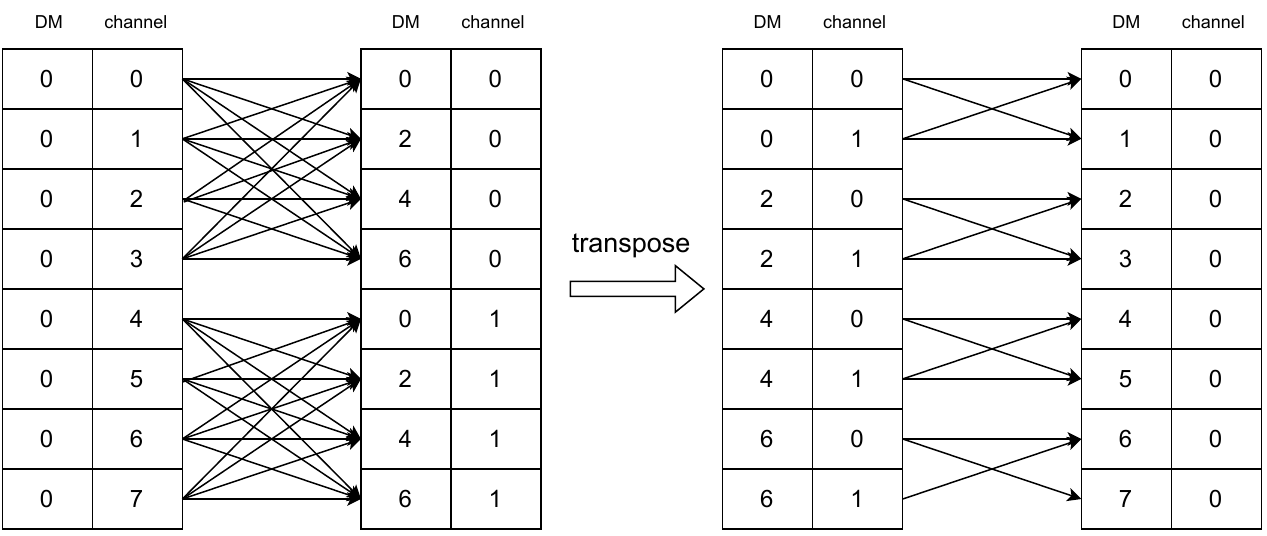}
    \caption{Example sub-band de-dispersion diagram with eight channels and DM trials. It presents the two stages of the sub-band de-dispersion: (1) de-dispersing the data into sub-bands with a coarser DM grid; (2) de-dispersing the sub-banded data with a finer DM grid.}
    \label{fig:subband}
\end{figure*}

To mitigate the S/N loss caused by DM mismatches, de-dispersion with DM trials is implemented in the single-pulse search. A straightforward approach to de-dispersion involves the brute-force algorithm, which calculates DM trials over a fine grid with a fixed DM step. This approach has a complexity of $\mathcal{O}(N_\RM{DM} N_t N_f)$, where $N_\RM{DM}$, $N_t$, and $N_f$ represent the number of DM trials, samples, and frequency channels, respectively. However, more efficient algorithms have been proposed with significantly lower complexity. Examples include the tree de-dispersion algorithm \citep{Taylor1974A&AS} and the fast dispersion measure transform (FDMT) algorithm \citep{Zackay2017ApJ}, both of which have a complexity of $\mathcal{O}(N_\RM{DM} N_t \log_2 N_f)$, as well as the sub-band de-dispersion \citep{Magro2011MNRAS} with a complexity of $\mathcal{O}(N_\RM{DM} N_t \sqrt{N_f})$. These algorithms are based on the concept that the S/N loss due to a DM mismatch is related to the frequency bandwidth, as shown in \EQ{zeta}. De-dispersion can be performed initially in sub-bands with a coarser DM grid, and then in sub-banded data with a finer DM grid. In {\sc{TransientX}}, we provide an efficient implementation of the sub-band de-dispersion algorithm. The speed of this sub-band de-dispersion algorithm can be comparable to the FDMT algorithm, even with a larger complexity, because it is more CPU-cache friendly and can utilize the L3-cache of modern CPU architectures effectively \citep{Men2023arXiv}. \FIG{dedispersion_flow} illustrates the dataflow diagram, comprising four steps: (1) Reading the data block by block and storing it in a buffer; (2) Transposing the buffer between the time and frequency dimensions to optimise the CPU-cache-friendly memory layout for efficient de-dispersion; (3) Performing sub-band de-dispersion on the transposed buffer and saving the de-dispersed data; (4) Shifting the data that has not been de-dispersed to the front of the buffer. The sub-band de-dispersion process includes three steps: (1) De-dispersing the data in multiple sub-bands using a coarser DM grid; (2) Transposing the sub-banded data between the DM and frequency dimensions; (3) De-dispersing the transposed sub-banded data into time series with a finer DM grid. An example of the sub-band de-dispersion process and the memory layout is depicted in \FIG{subband}, and the benchmark results are presented in \SEC{Benchmark}. Additionally, since the intra-channel smearing surpasses the native time resolution at high DMs, {\sc{TransientX}} supports de-dispersion plans, allowing for downsampling before de-dispersion in multiple DM ranges during a single processing step, further enhancing performance.

\subsection{Matched filtering}
\label{sec:Matched filter}

\begin{figure}
    \centering
    \includegraphics[width=\columnwidth]{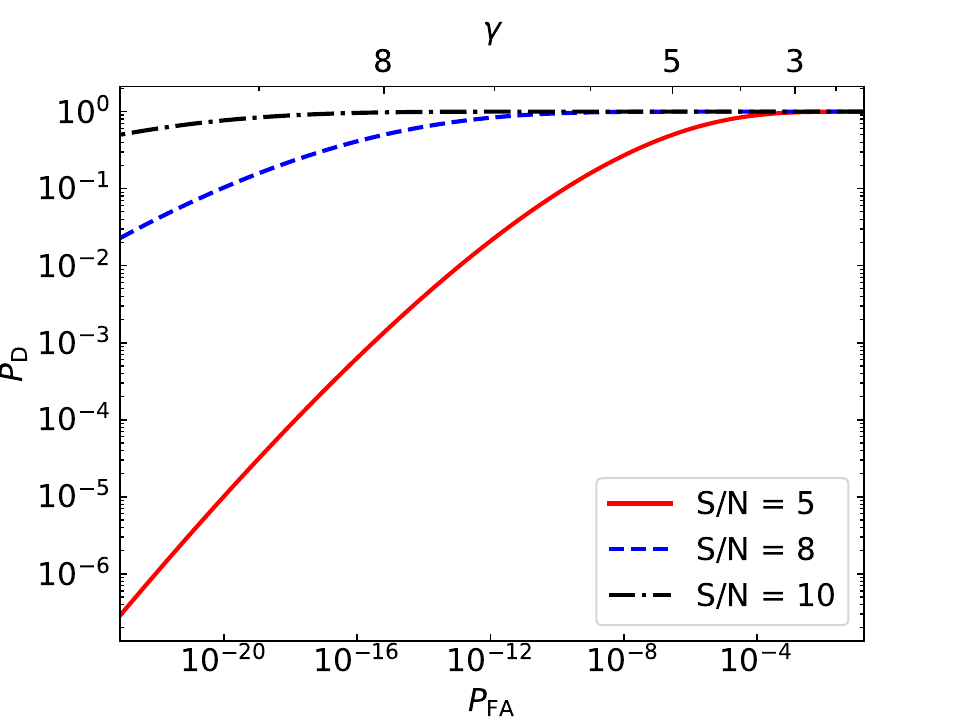}
    \caption{ROC curve. The red solid, blue dashed, and black dash-dotted lines represent the ROC curve for pulses with S/N values of five, eight, and ten, respectively. $\gamma$ is the S/N threshold.}
    \label{fig:roc_curve}
\end{figure}

\begin{figure}
    \centering
    \includegraphics[width=\columnwidth]{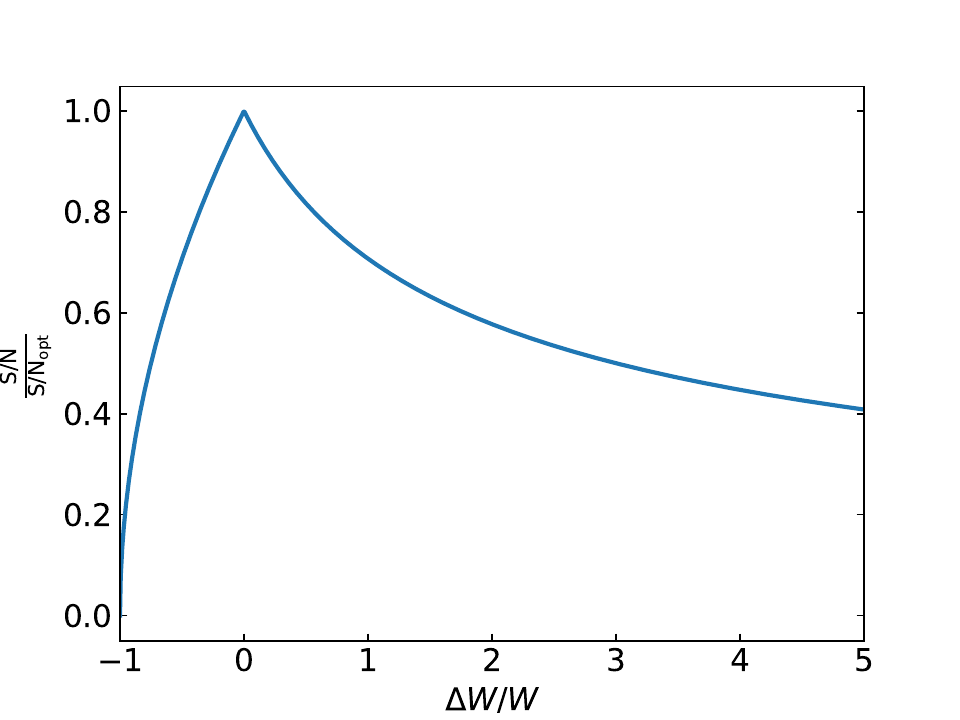}
    \caption{Decrease in S/N as a function of the boxcar width mismatch.}
    \label{fig:width_mismatch}
\end{figure}

Following the de-dispersion process, we obtain multiple time series corresponding to different DM trials. In accordance with the Neyman-Pearson lemma \citep{NeymanPearson1933}, the most efficient test for two hypotheses involves employing the maximum likelihood ratio test. Consequently, we utilize the matched filter algorithm to carry out pulse detection. To ensure efficiency in our implementation, we make an assumption that the pulse has a boxcar shape. We then apply a running boxcar window to the time series, calculating the S/N by
\begin{equation}
    \RM{S/N} = \frac{1}{\sqrt{N_\RM{box}} \sigma} \sum_{|t-t_0|<W/2} x(t)\,,
    \label{eq:snr}
\end{equation}
where $W$ represents the boxcar width and $t_0$ is the central time of the boxcar, $N_\RM{box}$ is the number of samples within the boxcar, and $\sigma$ is the standard deviation of the noise. The summation term in \EQ{snr} can be computed by two steps: (1) calculating the accumulation of the time series; (2) subtracting the accumulation at the end sample and the start sample of the boxcar window. This computation has a complexity of $\mathcal{O}(N_t)$ for one time series and one boxcar width. Under the assumption of Gaussian white noise, the probability of detection and false alarm for the statistics S/N can be expressed as
\begin{equation}
    P_\RM{FA} = \frac{1}{2} \RM{erfc}\left( \frac{\gamma}{\sqrt{2}} \right) \,,
\end{equation}
\begin{equation}
    P_\RM{D} = \frac{1}{2} \RM{erfc}\left( \frac{\gamma-\RM{S/N}}{\sqrt{2}} \right) \,,
\end{equation}
where $\gamma$ is the S/N threshold. \FIG{roc_curve} displays the receiver operating characteristic (ROC) curve. Given the unknown pulse width, we employ a running boxcar with various widths. We can deduce that the S/N loss resulting from a mismatch in the boxcar width, denoted as $\Delta W$, is 
\begin{equation}
    \frac{\RM{S/N}}{\RM{S/N_{opt}}} =
        \begin{cases}
            \sqrt{\frac{W+\Delta W}{W}} & \Delta W \le 0 \,, \\
            \sqrt{\frac{W}{W+\Delta W}} & \Delta W > 0 \,,
        \end{cases}
\end{equation} 
$\RM{S/N_{opt}}$ represents the optimal S/N achieved with the optimal width. \FIG{width_mismatch} illustrates this relationship. The extent of S/N loss relies on the ratio between the width mismatch and the true width. Hence, we utilize boxcar widths arranged in a geometric sequence with a tunable ratio to manage the S/N loss in {\sc{TransientX}}.

\begin{figure*}
    \centering
    \includegraphics[width=0.8\columnwidth]{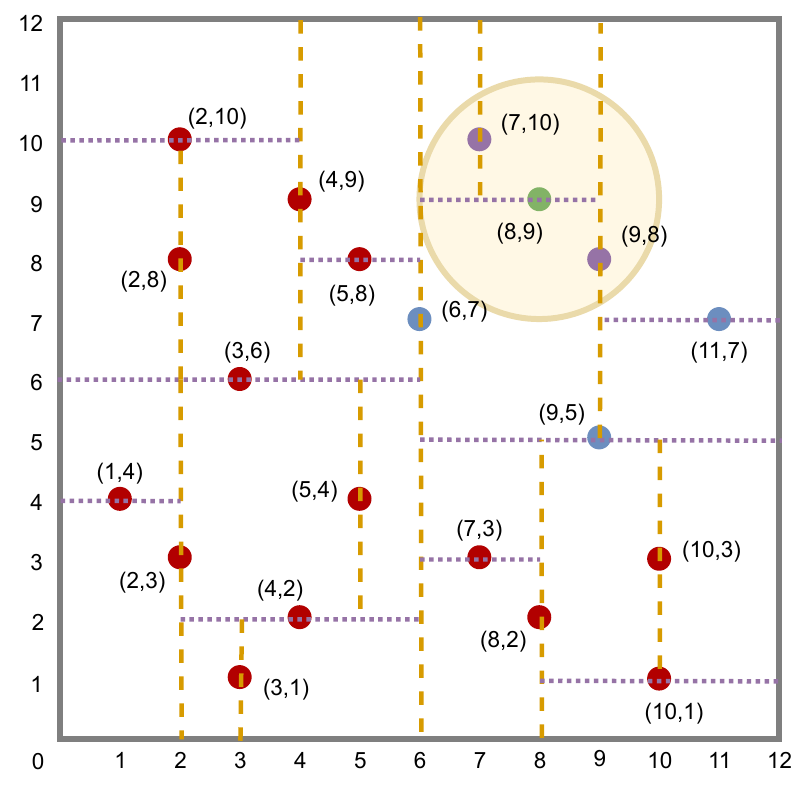}
    \includegraphics[width=0.8\columnwidth]{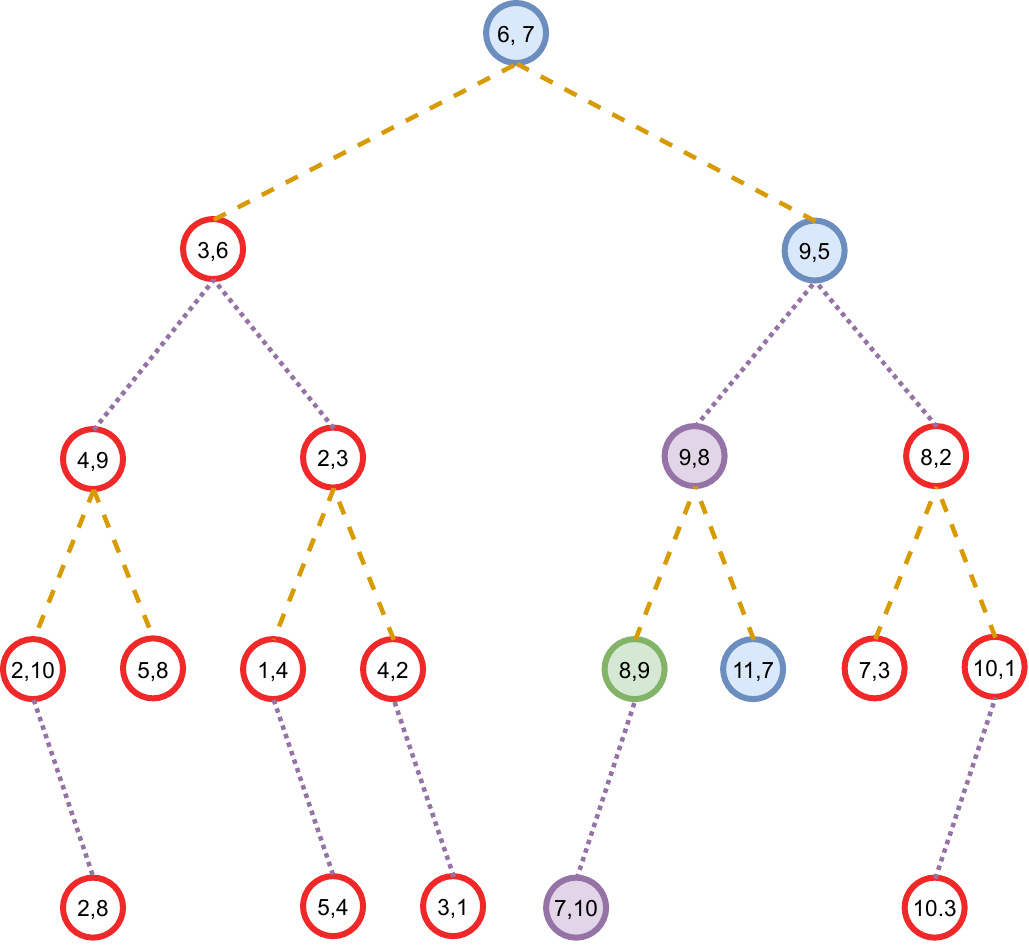}
    \caption{Diagram illustrates the neighbour-finding algorithm using a kd-tree. The example demonstrates the process of finding neighbour points within a radius of $r=2$ around a point (8, 9), denoted by a green circle. The left panel displays the plane divisions created by the points on the kd-tree. Yellow dashed lines and purple dotted lines divide the plane vertically and horizontally, respectively. The yellow shaded area represents the region within a radius of $r=2$ from the point (8, 9). The right panel illustrates the kd-tree structure employed to identify neighbour points around the point (8, 9). Red circles indicate points that won't be traversed, while others represent points that are traversed. The purple points highlight the neighbour points found within the radius of $r=2$.}
    \label{fig:clustering}
\end{figure*}

\subsection{Clustering}
\label{sec:clustering}
Following the pulse detection process, we obtain an S/N cube that encompasses DM, time, and width parameters. It's expected that if a pulse is located at a specific point within the S/N cube, we may observe neighbouring regions around the true parameters where the S/N values exceed the detection threshold. This can lead to the generation of multiple duplicate candidates for a single pulse, which is undesirable. To address this issue and eliminate these duplicate candidates, we employ the density-based spatial clustering of applications with noise (DBSCAN) algorithm \citep{Ester1996}. To enhance its performance, we initially compress the S/N cube into a two-dimensional DM-time plane. We achieve this by selecting the width with the highest S/N value for each DM and time combination. Additionally, we convert the DM dimension to dispersion delay using \EQ{dispersion_delay} to ensure that both dimensions share the same unit, such as milliseconds. Furthermore, we filter out points with S/N values below the pre-defined S/N threshold.

In the DBSCAN algorithm, several key notations are utilised:
\begin{itemize}
    \item \emph{core point}: A point is considered a core point if it has a number of neighbouring points that exceeds a pre-defined threshold $k$ within a specified radius $r$.
    \item \emph{reachable point}: A point is categorised as a reachable point if it is not a core point but lies within the radius $r$ of a core point.
    \item \emph{outlier}: A point is classified as an outlier if it neither qualifies as a core point nor a reachable point.
    \item \emph{density reachable}: Two points, denoted as $p_0$ and $p_n$, are regarded as density reachable if there exists a chain of core points, including $p_1, p_2, p_3, \dots, p_{n-1}$, where each adjacent point is within the radius $r$. Furthermore, both $p_0$ and $p_n$ must be within the radius of $p_1$ and $p_{n-1}$, respectively.
\end{itemize}
The algorithm can be summarised in the following steps: (1) choosing an initial point $p$; (2) finding out all density reachable points with $p$, which form one cluster if $p$ is a core point; (3) choosing another point if $p$ is not a core point; (4) iterating through steps (1), (2), and (3) until all points have been processed.

In {\sc{TransientX}}, we have developed an efficient C++ implementation of the DBSCAN algorithm. We leverage a space-partitioning data structure called a kd-tree to divide the DM-time plane. This approach provides an efficient means to locate points within a given radius, with a complexity of $\mathcal{O}(\log_2 N)$. \FIG{clustering} illustrates how the kd-tree partitions a two-dimensional plane. To find the neighbouring points of a given point $p_0$ within a radius $r$, the algorithm traverses the kd-tree. If the point $p$ on a node falls within the radius $r$, it is marked as a neighbouring point of $p_0$, otherwise, it is not considered. To enhance efficiency, branches can be eliminated from the traversal based on the following criteria: (1) If the point $p$ on the node serves as a vertical divider, and the horizontal distance between $p$ and $p_0$ exceeds the radius $r$, then the branch in the opposite half of point $p$ can be discarded; (2) If the point $p$ on the node acts as a horizontal divider, and the vertical distance between $p$ and $p_0$ surpasses the radius $r$, then the branch in the opposite half of point $p$ can be omitted. This neighbour point-finding algorithm greatly enhances the efficiency of the DBSCAN algorithm when applied in {\sc{TransientX}}.

\begin{figure}
    \centering
    \includegraphics[width=\columnwidth]{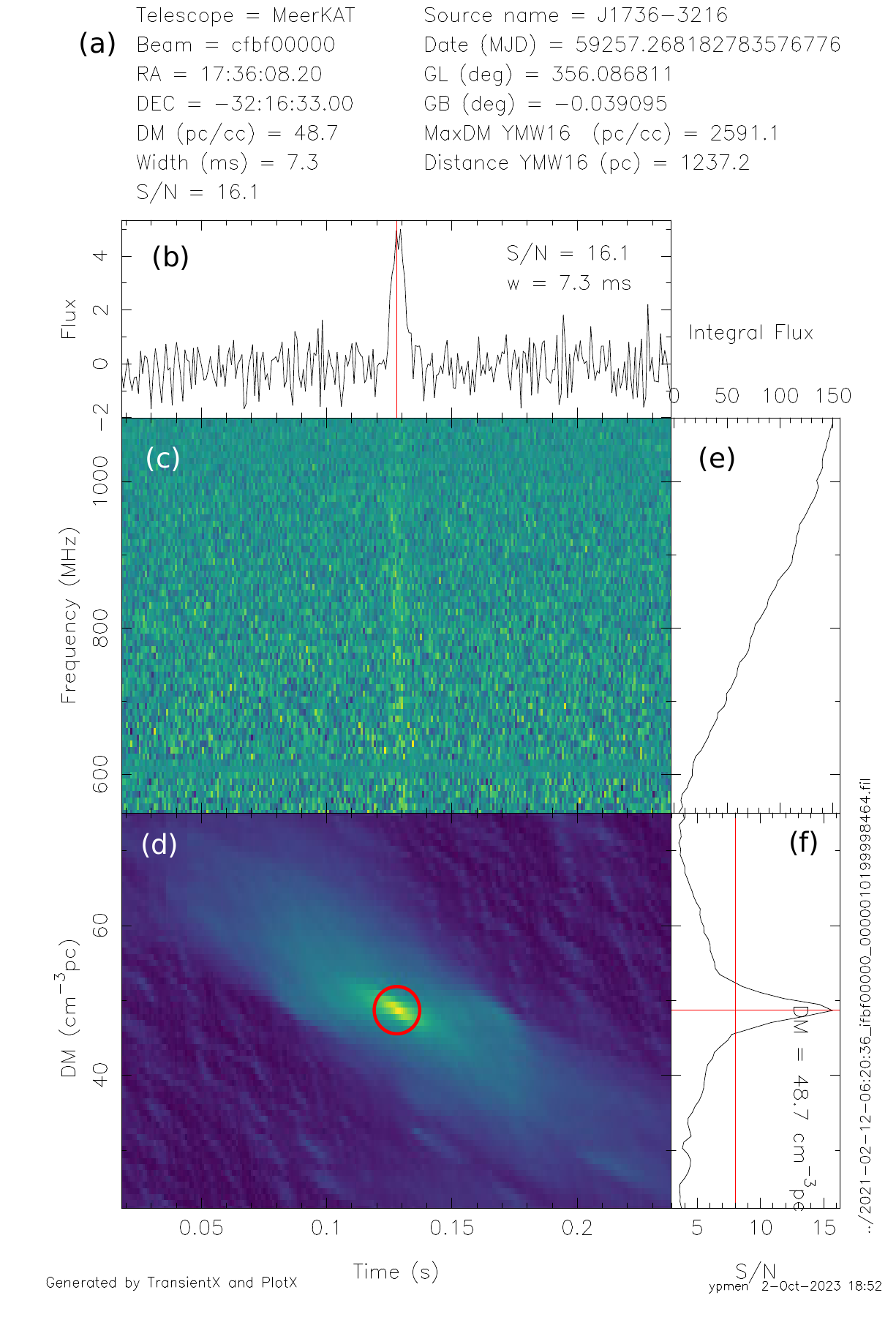}
    \caption{Example candidate plot, created by {\sc{TransientX}}. Panel (a) presents essential meta-information about the pulse. Panel (b) displays the pulse profile. Panel (c) depicts the pulse's dynamic spectrum. Panel (d) illustrates the S/N distribution in the DM-time plane after applying the matched filter. Panel (e) shows the pulse's bandpass. Panel (f) shows the S/N versus DM relationship. The vertical red solid lines represent the S/N threshold, and the horizontal solid line denotes the measured DM of the pulse. The data path of the pulse is indicated along the right border.}
    \label{fig:candidate}
\end{figure}

\subsection{Candidate plotting}
\label{sec:Candidate plotting}

After the removal of duplicate candidates in the clustering process, a significant number of candidates, including both genuine pulses and RFI signals, still remain. While machine learning-based tools like FETCH \citep{Agarwal2020MNRAS} can aid in candidate classification, visual inspection remains a crucial step. To facilitate this, we generate figures for each candidate, incorporating essential meta-information such as DM, time-of-arrival, boxcar width, and the maximum Galactic DM contribution in the direction of the source, based on the YMW16 model \citep{Yao2017ApJ}. Additionally, each figure includes dynamic spectra of the pulse and an S/N map in the DM-time plane, as exemplified in \FIG{candidate}. To streamline the process of creating these candidate figures efficiently, we've implemented a C++-based plotting library called {\sc{PlotX}\footnote{https://github.com/ypmen/PlotX}}. This library serves as a {\sc{matplotlib}}-like wrapper for the high-performance plotting library {\sc{pgplot}\footnote{https://sites.astro.caltech.edu/~tjp/pgplot/}}.

\section{Benchmark}
\label{sec:Benchmark}

{\sc{TransientX}} operates with a single command line, and the total elapsed time can vary even with the same configuration. This variability is due to the clustering process's dependence on the number of clusters and the number of points within those clusters, which can differ significantly. To provide a clear understanding of performance, we offer benchmarks for individual processes, encompassing de-dispersion, matched filtering, and clustering: (1) For the de-dispersion benchmark, we simulate datasets with varying numbers of frequency channels, resulting in different time costs. Additionally, we maintain the same number of DM trials as frequency channels. (2) For the matched filtering benchmark, since the execution time scales linearly with the number of DMs, we provide time costs for different boxcar widths, offering insights into the impact of this parameter. (3) For the clustering benchmark, as {\sc{TransientX}} processes data in small blocks, typically one second in duration, we calculate time costs for a single data block under different parameter configurations, specifically varying values of $r$ and $k$ in the DBSCAN algorithm. The data block contains a simulated single pulse, and we vary the number of points in a cluster by adjusting the S/N threshold.

\begin{figure}
    \centering
    \includegraphics[width=\columnwidth]{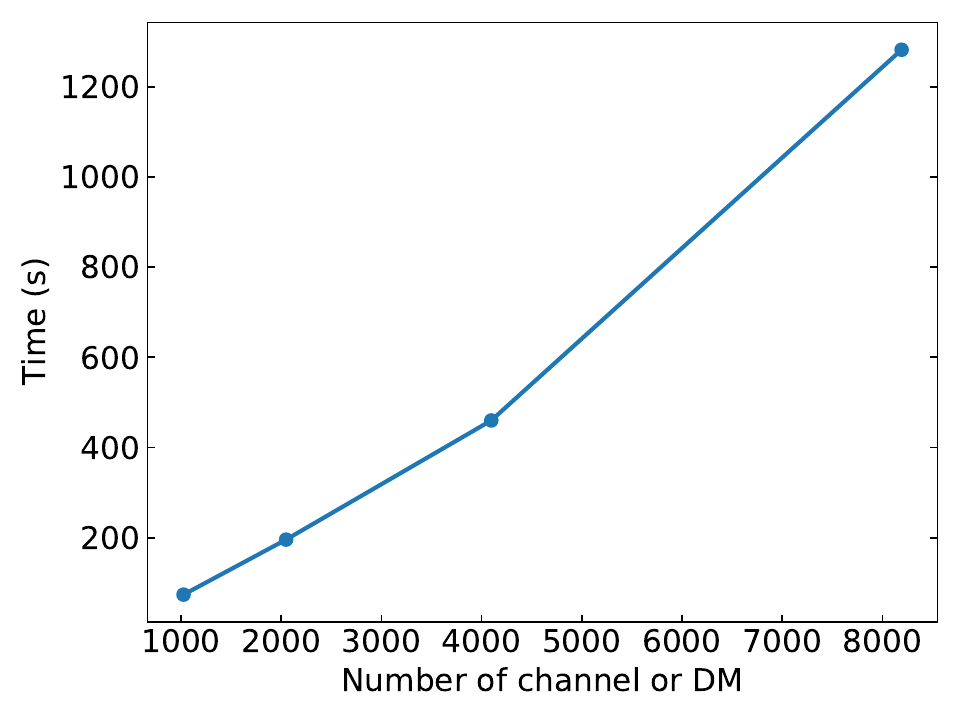}
    \includegraphics[width=\columnwidth]{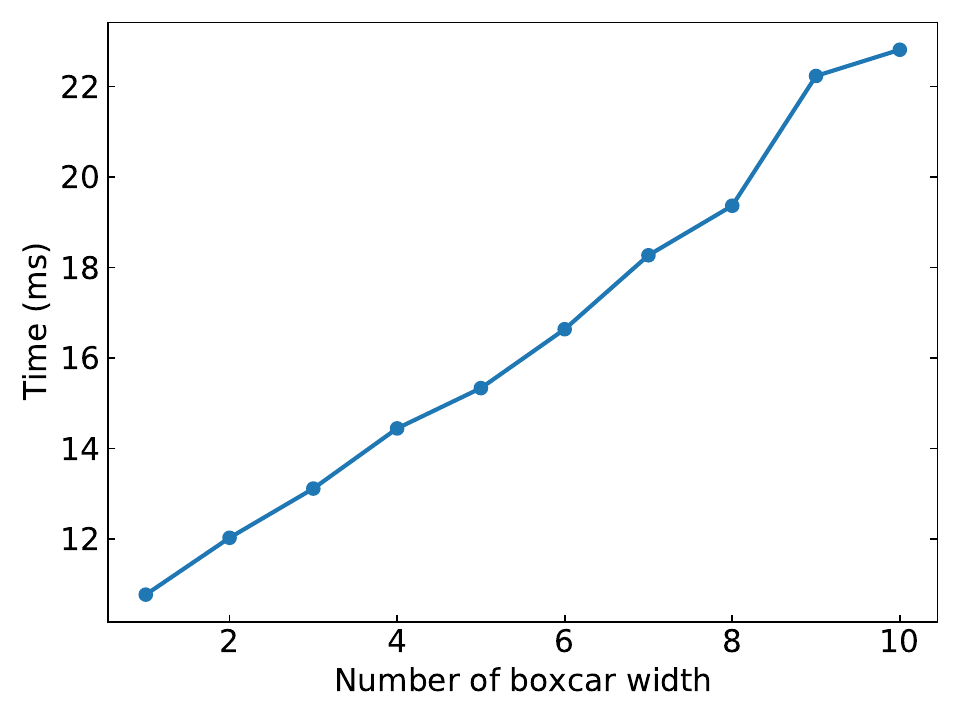}
    \includegraphics[width=\columnwidth]{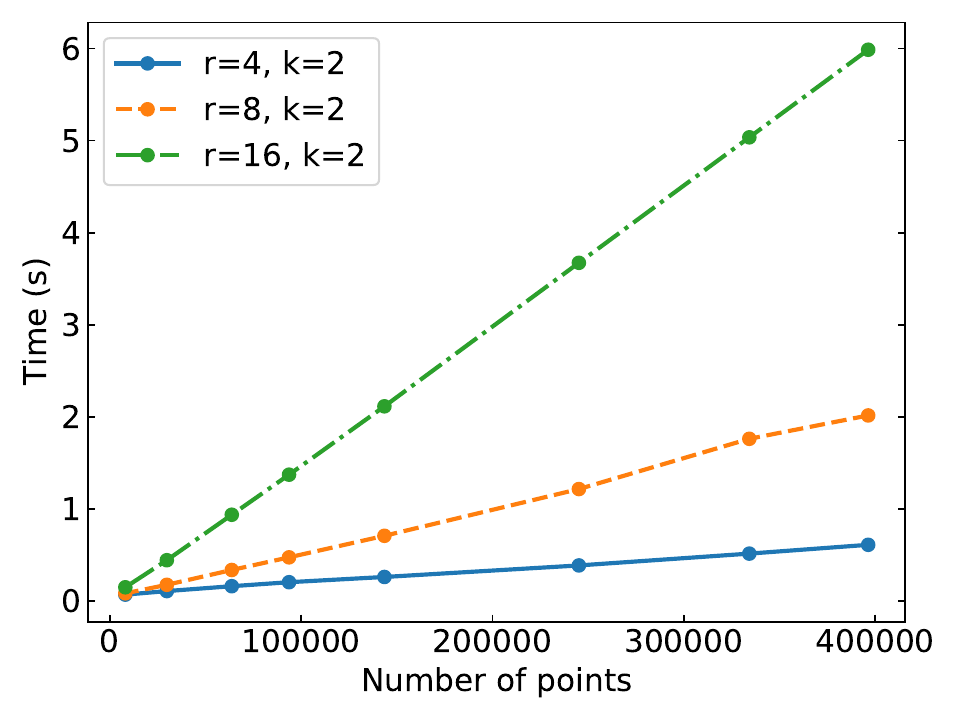}
    \caption{Benchmark results for de-dispersion, matched filtering, and clustering. The top-left panel depicts the relationship between the de-dispersion time cost and the number of frequency channels, which is equivalent to the number of DM trials. It demonstrates the efficiency of the de-dispersion process. The top-right panel shows the relationship between the matched filter time cost per DM trial and the number of boxcar widths. The linear scalability of the matched filter process with respect to the number of boxcar widths is evident. The bottom panel illustrates how clustering time scales with the number of points and is affected by the choice of radius, where the blue solid line, orange dashed line and green dash dotted line represent the radius $r$ of four, eight and sixteen, in the unit of samples, respectively.}
    \label{fig:benchmark}
\end{figure}

For the benchmarking process, we utilised a simulated dataset with a time resolution of 153 microseconds and a length of 600 seconds. This configuration is similar to the MPIfR-MeerKAT Galactic Plane Survey of L-band (MMGPS-L) \citep[MMGPS-L;][]{Padmanabh2023MNRAS, Bernadich2023arXiv}. The benchmark is conducted on a CPU model of Intel(R) Core(TM) i7-10750H. The results are presented in \FIG{benchmark}. From these results, several key observations can be made: (1) The time elapsed on the de-dispersion process, which accounts for a significant portion of the total time, is less than the data length of 600 seconds even with 4096 frequency channels. This high efficiency demonstrates the capability of {\sc{TransientX}} to handle large datasets. (2) The time elapsed on the matched filter process scales linearly with the number of boxcar widths. This linear scalability arises because the computational complexity is $\mathcal{O}(N_t)$ and does not depend on the boxcar widths, as discussed in \SEC{Matched filter}. (3) The time elapsed on the clustering process linearly scales with the total number of points to be clustered. Additionally, the clustering time increases with a larger radius $r$ in the unit of samples due to additional iterations in traversing the kd-tree. Concerning the entire single pulse search process, {\sc{TransientX}} demonstrates the capability to process MMGPS data with 2048 frequency channels and a 153-microsecond time resolution in real-time using a single CPU core, showcasing its efficiency in handling demanding radio astronomy datasets.

While our benchmark focuses on MMGPS configuration test data, {\sc{TransientX}} is adaptable to various data configurations, including low-frequency observations and microsecond timescales. In the pursuit of single-pulse detection in the lower radio frequency band (e.g. 100-200\,MHz), dispersion delay can exceed 10 seconds. However, due to scattering effects that broaden the pulse, downsampling to lower time resolutions can reduce memory and computing requirements. Nevertheless, it's important to note that sub-band de-dispersion might slow down if the data within sub-bands exceed the CPU L3-cache capacity. For microsecond burst searches, adjusting the data block length to fit the cache could be advantageous. In all cases, proper DBSCAN parameter configuration based on the time resolution during the clustering process is vital to prevent slowdowns.

\section{Discussions}
\cite{CHIME2018ApJ} mentioned that they searched for spectral index and scattering timescale, in addition to DM, pulse width, and time of arrival. As a potential future improvement for {\sc{TransientX}}, we have investigated the S/N loss caused by mismatches in spectral index and pulse shape.
\label{sec:Discussions}
\subsection{Mismatch of the spectra index}
In the de-dispersion process, the frequency channels are typically combined with equal weights, although this may not represent the optimal weighting scheme. This is because real astronomical signals often exhibit non-flat spectra. For instance, pulsar signals frequently follow a power-law with a negative index \citep{Bates2013MNRAS}. Hence, the ideal approach involves determining the spectral index as well. The spectral data $\pmb{x}$ with a spectral index $\alpha$ can be characterised as
\begin{align}
    x_{j,i} &= a_j s_i + n_{j,i}\,,\\
    a_j &= a_0 \left(\frac{f_j}{f_0}\right)^{-\alpha}\,,\\
    \left<n_{j,i} n_{j',i'}\right> &= \delta_{i,i'} \delta_{j,j'} \sigma^2\,,\\
\end{align}
where $f_j$ represents the frequency in the $j$-th channel, and $\sigma$ represents the standard deviation of the noise. The S/N loss resulting from a mismatch between the spectral index $\alpha$ and 0 can be expressed as
\begin{align}
    \frac{\RM{S/N}}{\RM{S/N_{opt}}} &= \frac{\left(\eta^{-\alpha+1}-1\right)}{-\alpha+1} / \sqrt{\frac{\left(\eta-1\right) \left(\eta^{-2\alpha+1}-1\right)}{-2 \alpha+1}} \,,\\
    \eta &= \frac{f_h}{f_l} \,,
\end{align}
where $f_h$ and $f_l$ are the highest and lowest frequency, respectively. The S/N loss caused by the mismatch of spectral index for both $\eta=2$ and $\eta=5$ is depicted in the left panel of \FIG{spectra_index}. From the results, we can observe that for a typical pulsar spectral index of 2, the S/N loss is only about 6\% when $\eta=2$. However, FRB spectra can sometimes be narrowband \citep{Kumar2021MNRAS}, leading to much larger S/N losses, as seen in the case with a larger spectral index. Moreover, the S/N loss becomes more pronounced as $\eta$ increases, particularly in the case of ultra-wide bandwidth receivers \citep[e.g.][]{Hobbs2020PASA}.

\subsection{Mismatch of pulse shape}
\begin{figure*}
    \centering
    \includegraphics[width=0.98\columnwidth]{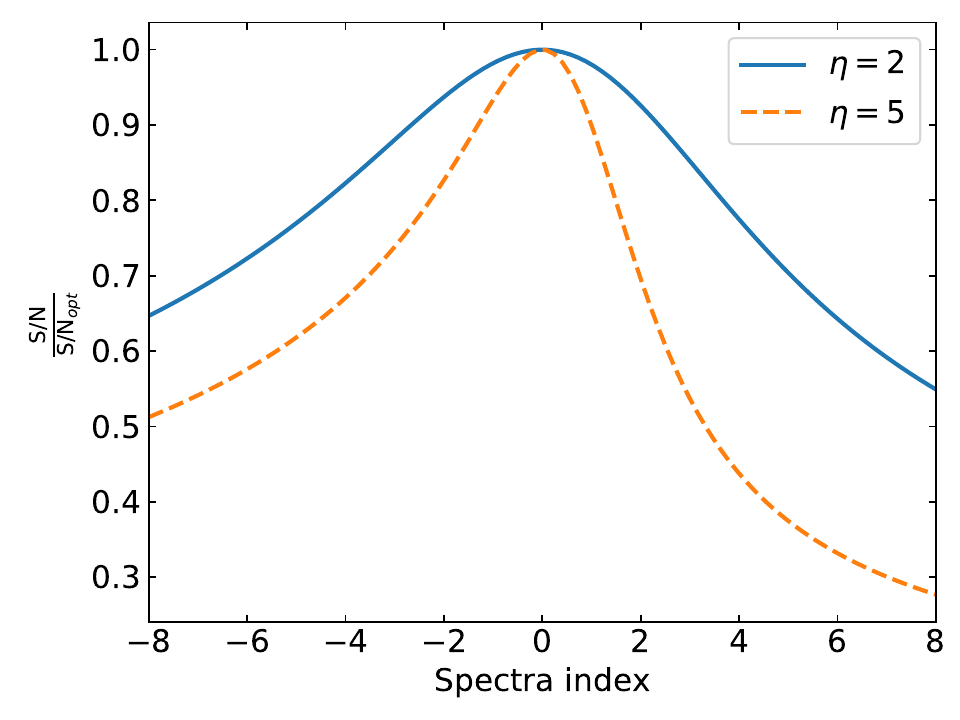}
    \includegraphics[width=\columnwidth]{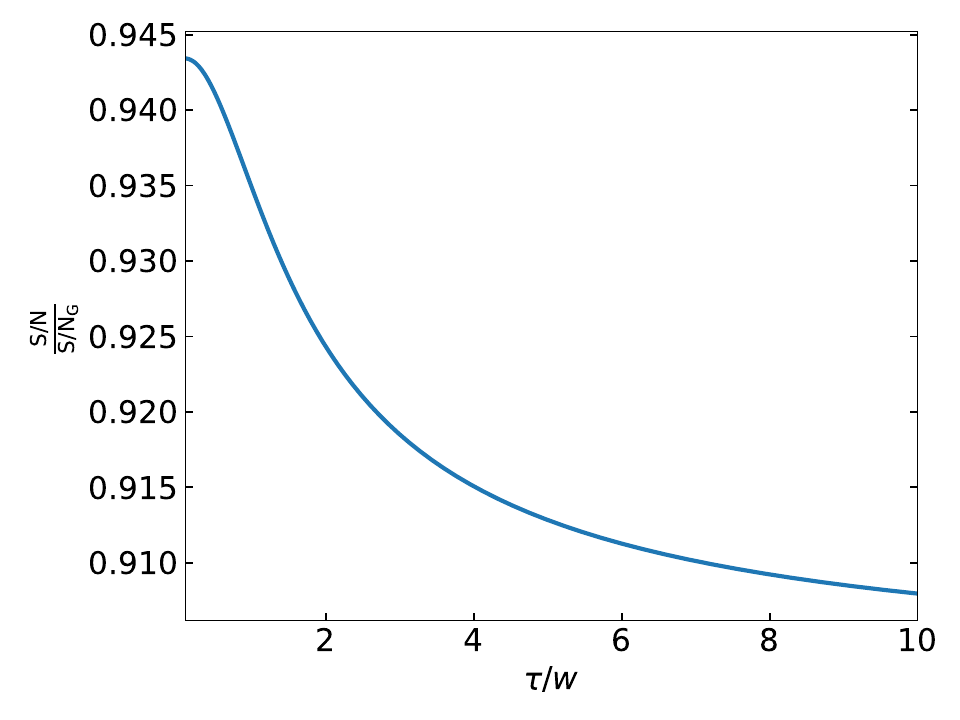}
    \caption{Signal-to-noise ratio decrease curves. The left panel illustrates the S/N decrease with a mismatch in the spectral index when $\eta=2$ and $\eta=5$. The right panel depicts the S/N decrease with a mismatch in pulse shape between a boxcar shape and a Gaussian shape with scattering.}
    \label{fig:spectra_index}
\end{figure*}

In the matched filter process, we employ a boxcar template, which may not accurately represent the true pulse shape and can result in S/N loss. To quantify this S/N loss, we derive a generalised S/N with a random pulse shape using the maximum likelihood ratio test,
\begin{equation}
\RM{S/N_G} = \frac{1}{\sigma} \frac{\sum_i x_i s_i}{\sqrt{\sum_i s_i^2}},
\label{eq:g_snr}
\end{equation}
where $s$ represents the pulse template. This equation reduces to \EQ{snr} when using the boxcar template. However, for the purpose of calculating the S/N loss, we consider a Gaussian-shaped pulse template, $s(t)$, with a scattering tail, given as
\begin{equation}
    s(t) = \exp \left(-\frac{t^2}{w^2}\right) \ast \left(\frac{1}{\tau_s}\exp\left(-\frac{t}{\tau_s}\right)\right)\,,
\end{equation}
where $\ast$ represents the convolution. $w$ and $\tau_s$ are the pulse width and scattering timescale, respectively. We then calculate the ratio between the S/N defined in \EQ{snr} and \EQ{g_snr}, and the results are shown in the right panel of \FIG{spectra_index}. The S/N loss caused by the mismatch of pulse shape between a boxcar shape and a Gaussian shape with scattering is small, approximately 9\%, even when the scattering timescale is 10 times the pulse width.

\section{Conclusions}
\label{sec:Conclusions}
In this study, we introduced a new high-performance single pulse search software, {\sc{TransientX}},  designed to be user-friendly and capable of running with a single command line while generating candidate figures. {\sc{TransientX}} features an efficient CPU-based implementation of the sub-band de-dispersion algorithm and an efficient implementation of the DBSCAN algorithm. These optimisations enable {\sc{TransientX}} to provide efficient CPU-only real-time single pulse searching capabilities. The application of {\sc{TransientX}} in the Transients and Pulsars with MeerKAT (TRAPUM) project will be presented in future work (Carli et al., in preparation). Additionally, we conducted analyses to quantify the S/N loss resulting from discrepancies in DM, pulse width, spectral index and pulse shape. These investigations have improved our understanding of S/N losses in single pulse searches: (1) The S/N decrease due to the DM mismatch exhibits a slower decline compared to the amplitude decrease; (2) The S/N decrease resulting from the spectra index mismatch becomes increasingly significant with multi-octave receivers; (3) The S/N decrease caused by the mismatch of pulse shape between a boxcar and a Gaussian shape with scattering remains minor. It is worth noting that in a survey, the rate of observed burst events might decline more rapidly than the decrease in S/N. This can be attributed to a steep S/N distribution, such as a power-law distribution with an index of -2, characterizing a uniform distribution of FRBs in the universe. As we approach the era of the Square Kilometre Array (SKA), which will produce vast amounts of data, the need for real-time data processing becomes increasingly critical. Advanced hardware and software solutions are essential to handle such data volumes effectively. The experiments on {\sc{TransientX}} and findings presented in this work can serve as valuable guidance for the development of future single pulse search pipelines, especially for SKA data processing.

\begin{acknowledgements}
The MeerKAT telescope is operated by the South African Radio Astronomy Observatory, which is a facility of the National Research Foundation, an agency of the Department of Science and Innovation. SARAO acknowledges the ongoing advice and calibration of GPS systems by the National Metrology Institute of South Africa (NMISA) and the time space reference systems department of the Paris Observatory.

TRAPUM observations used the FBFUSE and APSUSE  computing clusters for data acquisition, storage and analysis. These clusters were funded and installed by the Max-Planck-Institut für Radioastronomie and the Max-Planck-Gesellschaft.

YPM and EB acknowledge continuing support from the Max Planck society. 
\end{acknowledgements}

\bibliographystyle{aa}
\bibliography{ms}

\begin{appendix}
\section{S/N decrease caused by a DM mismatch}
\label{sec:dm_mismatch}
\begin{figure}
    \centering
    \includegraphics[width=\columnwidth]{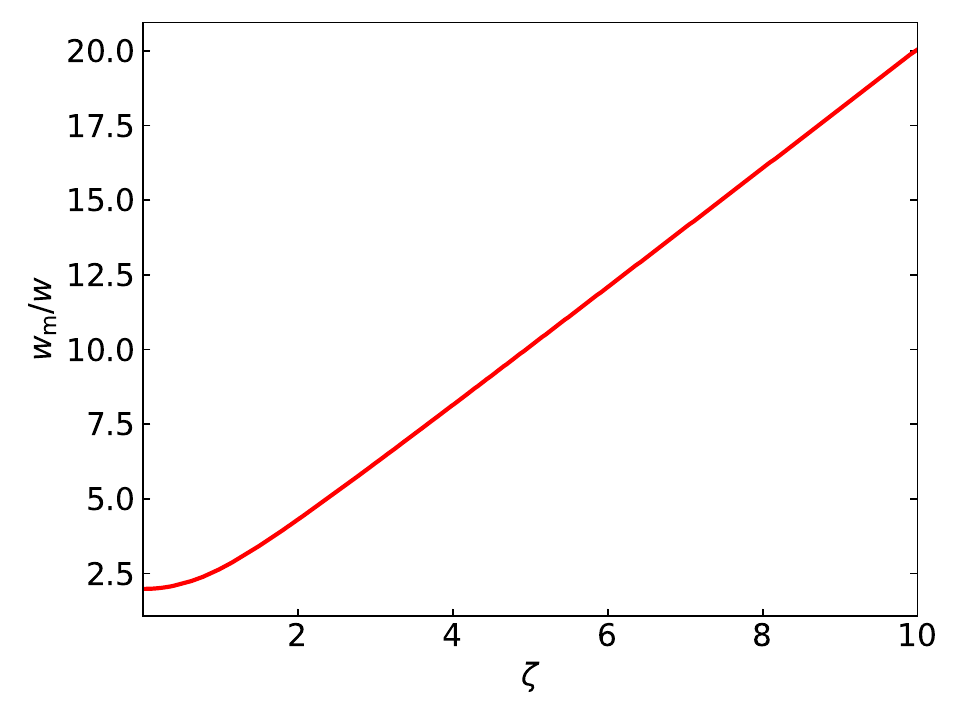}
    \caption{The optimal width relation with the DM mismatch factor $\zeta$ defined in \EQ{zeta} that maximizes the measured S/N in the matched filter.}
    \label{fig:w_m}
\end{figure}

For a radio pulse with a Gaussian profile, the dynamic spectrum is

\begin{align}
    x\left(f, t\right) &= \exp{\left(-\frac{(t - \tau(f))^2}{w^2}\right)} \label{eq:x_f_t}\\
    \tau(f) &\approx -2 \alpha \frac{\Delta \RM{DM}}{f_\RM{c}^3} (f - f_\RM{c}) \\
    \alpha &= \frac{e^2}{2\pi m_e c}
\end{align}
where $w$ is the pulse width and $f_\RM{c}$ is the central frequency. By integrating $x\left(f, t\right)$, we can get the de-dispersed profile,
\begin{equation}
    s\left(t\right) = \frac{\sqrt{\pi}}{4} \zeta^{-1} \left( \RM{erf}\left( \zeta - \frac{t}{w} \right)  + \RM{erf}\left( \zeta + \frac{t}{w} \right) \right) \,,
\end{equation}
where $\zeta$ is defined in \EQ{zeta}. From \EQ{snr}, we can get
\begin{align}
    \RM{S/N} &\propto \frac{1}{\sqrt{w_\RM{m}}} \int^{w_\RM{m}/2}_{-w_\RM{m}/2} s\left(t\right) d t \,,
\end{align}
where $w_m$ is the measured pulse width that gives the maximum S/N. Since $w_\RM{m}$ can't be derived analytically, we calculate it numerically, which is close to $2 \zeta w$ as $\zeta$ becomes larger, as shown in \FIG{w_m}. For a given $\zeta$, we can then calculate the S/N decrease, as shown in \FIG{dm_mismatch}.

\end{appendix}

\end{document}